\documentclass{article}
\usepackage[utf8]{inputenc}	
\usepackage[T2A]{fontenc} 
\usepackage{graphicx} 
\usepackage{setspace} 
\usepackage{amsmath, amsfonts}  
\usepackage{microtype}  
\usepackage{natbib}  
\usepackage[toc,page]{appendix}  
\usepackage{authblk} 
\usepackage{xurl} 
\usepackage{booktabs} 
\usepackage{bm} 

\usepackage{subcaption}  
\usepackage{array}
\usepackage{makecell}

\newcolumntype{C}{>{\centering\arraybackslash}p{1.55cm}}
\newcolumntype{L}{>{\raggedright\arraybackslash}p{2.45cm}}

\newcommand{\colhead}[1]{%
  \parbox[t][1.45cm][t]{1.55cm}{\centering #1}%
}

 \textwidth15.75cm \evensidemargin5mm \oddsidemargin5mm
\title{Monetary Regimes and Trade before the Classical Gold Standard: Evidence from the Latin Monetary Union\footnote{An earlier version of this paper circulated under the title ``The Latin Monetary Union and Trade: A Closer Look''. I am deeply grateful to Christopher M. Meissner for sharing his database on monetary standards. I thank Teodoro D'Agostino for excellent research assistance in coding RICardo bilateral trade flows to standardized ISO codes. I also thank Rodolfo G. Campos, Marc Badia-Miró, Federico Barbiellini Amidei, Matteo Gomellini, and Michele Mancini for helpful comments and suggestions, as well as other participants at the Eternal City Economic History Workshop ``Stefano Fenoaltea'' at Banca d'Italia, and the Banco de Espa\~na research seminar. The views expressed in this paper are those of the authors and do therefore not necessarily reflect those of the Banco de Espa\~na or the Eurosystem.}}

\author[]{Jacopo Timini}
\affil[]{\it Banco de España}

\date{\today}

\begin{document}

\maketitle
\begin{abstract} \noindent
This paper reexamines the trade effects of the Latin Monetary Union (LMU), a 19\textsuperscript{th} century agreement to standardize gold and silver coinage among several European countries.
The LMU provides a useful setting for studying whether monetary arrangements fostered trade before the classical gold standard, when gold, silver, bimetallic, and paper regimes coexisted.
Because some countries already shared other monetary standards, treating all non-member pairs as a single control group mixes pairs with and without alternative forms of monetary coordination. I classify pairs by standard and estimate the LMU effect relative to pairs without a common standard, bringing the comparison closer to those used in the literature on the gold standard and contemporary currency unions.
The results suggest that the LMU increased trade between its members by approximately 30\% during its early years, when bimetallism was still credible. These effects subsequently faded, converging to zero by the end of the 1870s. More broadly, these findings also highlight the importance of accounting for the existing monetary regimes when estimating the trade effects of other international policies.
\end{abstract}



\clearpage

\begin{doublespace}

\section{Introduction}

Do currency unions boost trade between their members? This question has driven hundreds of papers and generated more than 3000 estimates of the Euro's effect on trade alone \citep{Polak:19}. While most studies focus on the post-World War II period, a growing body of research has turned to earlier monetary arrangements. In this context, the present paper reexamines the Latin Monetary Union (LMU)---a 19\textsuperscript{th} century agreement among several European countries to standardize their currencies through a bimetallic system based on fixed gold and silver content—--and its effects on trade, by incorporating the latest advances in gravity modeling and explicitly accounting for the diverse landscape of European currency regimes during the early years of the LMU.

Although economic theory and historical accounts suggest several channels through which the LMU could have facilitated trade between its members, at least during its early years, previous studies have not found robust evidence of a positive trade effect for the LMU as a whole.
The key theoretical benefit of a contemporary currency union is that, by adopting a common currency, countries remove the need for currency conversion, thereby reducing transaction costs and uncertainty in cross-border pricing \citep{MiccoSteinOrdonez:03}. However, the LMU did not create a common currency in this modern sense. As emphasized by \citet{Flandreau:04}, exchange rate volatility among countries on metallic monetary regimes was relatively low by modern standards before the 1870s. Yet low exchange rate volatility does not imply that monetary arrangements were irrelevant for trade. By aligning the weight, fineness, and denomination of gold and silver coins, and by allowing coins issued by one member state to circulate and be accepted at face value in the others, the LMU could reduce expected trade costs through other margins: lower verification and settlement costs, greater certainty over the future maintenance of mint standards, and a reduced scope for unilateral debasements or restrictions on circulation. 
These frictions did not arise from an inherent instability of bimetallism. As shown by \citet{Flandreau:04}, French arbitrage between gold and silver helped stabilize their relative value at the official parity. However, this mechanism did not coordinate national decisions concerning coin specifications, subsidiary coinage, or the cross-border acceptance and repatriation of coins. The LMU introduced common rules in these areas. Its possible effect on trade therefore came from the stabilization of expectations about future monetary conduct and the credibility of continued monetary cooperation among member countries.

The LMU went beyond the \textit{de facto} alignment of its members’ bimetallic systems by establishing common coinage rules and mutual commitments. Its possible effect on trade therefore came not only from bimetallism itself, but also from making the existing arrangement more predictable and reducing the risk of unilateral changes in national coinage policies. Indeed, these standards had been repeatedly disrupted by fluctuations in the relative prices of gold and silver around the mid-19\textsuperscript{th} century, which had triggered recurrent unilateral—--and thus uncoordinated—--adjustments to national monetary systems \citep{Einaudi:00}. However, around the mid-1870s, the credibility of bimetallism collapsed \citep{FlandreauOosterlinck:12,Meissner:15}, marked by a steep drop in silver’s value. This undermined the very basis of the LMU: the fixed gold-to-silver ratio became untenable, forcing members to limit and later suspend free silver coinage (1873-1874). As investors were quickly pricing in the world predominance of the gold standard, the LMU eventually kept trying to adapt to a changing world. However, later reforms only further signaled its emerging weaknesses. Although the LMU officially ended in 1927, the Union had become dysfunctional much earlier \citep{Gillard:17}.

Previous studies suggest that the LMU’s effects on trade were generally negligible or modest and, where positive, concentrated in specific bilateral flows \citep{Flandreau:00, Timini:18}, a finding that contrasts with the consistently positive trade effects attributed to other pre-WWI fixed exchange-rate regimes, such as the gold standard \citep{LopezCordovaMeissner:03, EstevadeordalFrantzTaylor:03, MitchenerShizumeWeidenmier:10, Badia-MiroCamposTimini:25}. This contrast is informative because the gold standard, like the LMU, was not a modern currency union: countries retained national currencies, central banks, and political and financial sovereignty, and the system relied on credibility and convertibility rather than on a common monetary authority \citep{ORourkeTaylor:13}.
The relevant empirical question is therefore whether LMU membership increased trade relative to pairs without a shared monetary standard, after accounting for other common regimes. The appropriate control group is therefore not composed of all non-LMU pairs, but of pairs that did not share any monetary arrangement. This distinction is particularly important in the monetary setting of the 1860s, when gold, silver, bimetallic, and paper regimes coexisted. Non-LMU pairs sharing one of these arrangements may already have benefited from monetary coordination and therefore differ from pairs without a common regime in ways that matter for the estimated LMU effect.

This paper makes two contributions. First, it provides evidence of a positive trade effect for the LMU as a whole. This finding brings the LMU evidence closer to the broader literature on historical monetary arrangements, which generally finds positive trade effects for the pre-WWI gold standard. While the two regimes differed institutionally, both could facilitate trade by reducing some monetary frictions and increasing the predictability of cross-border transactions.
Second, it shows the importance of distinguishing among the monetary regimes represented in the control group. This provides a more appropriate control group for the LMU and offers a methodological lesson for other studies of periods in which several monetary regimes coexisted.

My findings suggest that the LMU increased trade between its members by approximately 30\%, in its early years (1865-1873), coinciding with the period when bimetallism was still considered a viable monetary system. These effects then started fading, rapidly converging to zero by the end of the 1870s. 
Results are robust across a range of specifications, including the inclusion of additional potential confounders, the use of various samples spanning different countries and trade data sources, and alternative methodological choices.

The paper proceeds as follows. Section 2 provides the historical context, describing the creation of the LMU, the monetary frictions it aimed to address, and the loss of credibility of bimetallism from the early 1870s onward. Section 3 presents the empirical strategy and the data, with particular attention to the coding of monetary standards and the construction of the control group. Section 4 reports the main results, traces the evolution of the LMU effect over time, and presents robustness tests. It also uses the estimated trade cost reductions in a general equilibrium gravity framework to quantify the implied trade gains and to assess how these gains differ across LMU members. Section 5 concludes.

\section{Historical context}

The LMU was established in 1865 by France, Belgium, Italy, and Switzerland, later joined by Greece in 1868. These countries agreed upon a ``monetary convention'' aimed at harmonizing their monetary legislation, thereby remedying the ``inconveniences'' caused by differing coinage systems in cross-border transactions, as stated in the convention itself. The practical relevance of these frictions is documented in Willis's historical account of the early 1860s, which notes that monetary heterogeneity could mean that ``frontier trade was impeded, and travelers were subjected to inconvenience'' \citep[p. 40]{Willis:01}. The presence of a trade-related motivation is also consistent with \citet{FendelMaurer:15}, who emphasize that the creation of the LMU responded both to the search for a more stable monetary framework and to the objective of facilitating trade among neighboring economies.

This trade channel should nevertheless be understood in historically specific terms. The LMU did not create a common currency or a central monetary authority, and it did not operate mainly by eliminating large exchange rate fluctuations among its members. Exchange rate volatility among countries on metallic standards was already relatively low before the 1870s \citep{Flandreau:04}. The relevant margin was therefore not the disappearance of large exchange rate movements, but the reduction of uncertainty surrounding coinage standards, cross-border acceptance, and future monetary conduct.

This distinction is important because the existence of cross-border coin circulation before 1865 did not make the LMU irrelevant. Such circulation rested on national legislation, reputation, and arbitrage, while individual governments retained discretion over minting rules and coinage standards. The problem was therefore not the instability of bimetallism itself, but the absence of formal coordination among countries whose coins already circulated across borders. Although French arbitrage helped stabilize the relative value of gold and silver, individual governments retained discretion over coin specifications, the issuance of subsidiary coins, and the conditions under which foreign coins were accepted or repatriated.
The LMU sought to address these problems by creating a common institutional basis. By codifying coinage standards and mutual obligations, it reduced the scope for uncoordinated national action compared with the pre-LMU period. The new framework did not prevent all departures from these commitments. However, such episodes did not necessarily lead to an immediate loss of confidence in the Union. As long as bimetallism remained credible, they may have been seen as temporary difficulties that could be managed within the existing framework.

This institutional role was particularly relevant in the monetary context of the mid-19\textsuperscript{th} century. The existing bimetallic system helped maintain the gold-silver ratio through arbitrage, but changes in the composition of coin circulation created practical problems for countries whose currencies were already closely aligned. Governments responded separately, especially in relation to subsidiary silver coinage, creating differences in fineness and acceptance across borders. The LMU sought to coordinate these decisions by establishing common rules for coin specifications, issuance, acceptance, and repatriation. Although it could not rule out national interventions, it may have made them less likely and strengthened expectations that member
countries would refrain from unilaterally debasing their coinage or altering the agreed standards, both of which had been persistent concerns in earlier years. In this way, it may have reduced uncertainty over the future acceptance ad circulation of member countries’ coins \citep{Einaudi:00}.

More precisely, the LMU required that gold and silver coins be minted according to uniform standards of weight, fineness, and denomination. The convention defined one monetary unit as either 4.5 grams of fine silver or 0.29 grams of fine gold, reflecting a fixed gold-to-silver ratio of 15.5:1. This arrangement harmonized the coinage systems of member states: gold coins and the highest silver denomination---the 5-franc piece---were minted with 90\% fineness, while smaller silver denominations were struck at 83.5\%.\footnote{The reduced silver content in smaller denominations was a deliberate policy to discourage arbitrage and the melting of coins for bullion, a common issue in bimetallic systems.}
All member states committed to adhering strictly to these standards in their coin production. In practice, this meant that the French franc, Belgian franc, Italian lira, and Swiss franc were equivalent in value and metal content, making them interchangeable.\footnote{While it standardized the weight, fineness, and denomination of gold and silver coins of its member states---allowing them to circulate freely at face value---it did not establish a central monetary authority or a unified monetary policy. Each member retained sovereignty over its own minting and fiscal decisions, which meant that monetary coordination was limited to coin specifications rather than broader economic governance. Therefore, despite its name, it was a coinage union rather than a full monetary union \citep{Einaudi:01}.}

Any LMU member's coins circulated freely in all other LMU countries, with public offices, banks, and individuals potentially accepting coins from other LMU members at face value. A French merchant, for instance, could accept payment in Italian lire, Belgian francs, or Swiss francs with greater confidence that those coins would hold the same value as domestic French francs and could be exchanged back home.

This ``interoperability'' was backed by each government's commitment to honor the others' money. Public offices, such as national treasuries and mints, were obliged to accept gold coins and large silver 5-franc pieces from any member country without discrimination, just as they would accept their own coinage.\footnote{Smaller denomination silver coins, which were essentially token coinage, were subject to some limits in cross-border circulation to prevent overflow of small change: foreign subsidiary coins had to be accepted by government treasuries only up to a fixed amount, often 100 francs in total per payment. Beyond that threshold, a treasury could refuse excess foreign small coins, encouraging their repatriation.}
Central banks---or other issuing institutions---in member countries, such as the Banque de France, also played an important role by absorbing inflows of coins while maintaining convertibility and managing specie reserves. In this way, the LMU architecture was designed to reduce transaction costs among its members, which is, in principle, expected to have a positive impact on trade.\footnote{Currency unions are generally expected to influence member economies through two main channels. The first is the trade channel, which operates via reduced transaction costs, exchange rate stability, and increased price transparency---factors that typically promote greater cross-border trade. The second is the financial channel, which involves deeper financial integration, improved capital mobility, and potentially lower risk premiums due to shared monetary frameworks. This paper focuses exclusively on the trade channel. \citet{BordoRockoff:96} and \citet{BaeBailey:11} analyze the financial channel for the gold standard and the LMU respectively.}

The arrangement worked relatively well in the early years. However, the LMU's stability depended on each member upholding the agreed standards and on external economic conditions. The main external challenge came from the changing international monetary environment in the early 1870s, when the market price of silver fell relative to gold.\footnote{\citet{FendelMaurer:15} list many reasons why silver depreciated with respect to gold, including an increase in silver supply, coming both from production and from the move away from silver of some major countries, such as Germany, as well as changing preferences in the use of different monetary standards. They also provide an in-depth description of the LMU institutional structure.}
As the fixed legal ratio between gold and silver remained unchanged, the legal ratio diverged from the market ratio. Silver became overvalued in legal terms: individuals could profitably bring silver bullion to the mint, have it coined into legal tender, and then exchange those coins for gold at face value. This arbitrage opportunity favored the minting and circulation of silver coins, while gold coins were hoarded, exported, or withdrawn from banks. The result was a growing dominance of silver in everyday transactions and a significant depletion of gold reserves.

It was these imbalances that first prompted France to suspend free silver coinage in late 1873, after which the other LMU countries followed suit in early 1874.  By introducing quotas on silver coinage, LMU members suspended a central element of the bimetallic mechanism, namely the free conversion of silver bullion into coin at the official parity \citep{FlandreauOosterlinck:12}. A few years later---in late 1876 in France, and in 1878 for all LMU countries---the minting of new silver coins was finally halted. The LMU, however, maintained the legal tender status of existing large silver coins already in circulation. This situation became known as the ``limping gold standard'' \citep{BordoJonung:00, FendelMaurer:15}. In this context, existing large silver coins became a liability, unequally distributed among LMU members, as France ``held much more coin issued from the Mints of Belgium and Switzerland, and to some extent Italy, than was held by these Governments of the French coins'' \citep[as cited in \citeauthor{Timini:18}, \citeyear{Timini:18}]{nyt1885}.

To address this issue, LMU members agreed to amend the original convention by introducing a ``liquidation clause'': in the event of the dissolution of the LMU, each member state was obligated to repurchase its own large silver coins held by other members at face value. \citet{Willis:01}, one of the most prominent historical analysts of the LMU, suggested that ``the ratification of the treaty of 1885 really meant the abrogation of the Latin Union''. From 1885 onward, there has been broad consensus on the limited relevance of the LMU, despite its formal dissolution only in 1927.

\section{Methodology and Data}
\subsection{Methodology}

My analysis of the LMU effects on trade relies on gravity trade theory. As it is well-known \citep{HeadMayer:14, YotovPiermartiniMonteiroLarch:16}, gravity theory predicts that trade flows between two countries depend on their economic size (relative to the world) and the existing trade costs between them, and with the rest of the world.
More formally, this can be written as follows:

\begin{equation} \label{eq: gravity theory}
    X_{ij} = \frac{Y_{i} E_{j}}{Y} (\frac{\tau_{ij}}{\Omega_{i}\Pi_{j}})^{-\theta}
\end{equation}

$X_{ij} \ge 0$ denote international trade flows from country $i$ (the exporter) to country $j$ (the importer).\footnote{While ideally I would like to include both domestic and international trade data in my main specification, I have to face important data restrictions for at least one country (Switzerland), out of five LMU members: there is no sufficient data to compute domestic trade flows during the early years of the Union. Given the focus of the paper on the evolution through time of the LMU trade effects, and the few countries that composed the Union, this can introduce a bias. Therefore, I prefer to adopt a conservative approach and estimate my main specification with international trade only, while later checking the robustness of the results to the inclusion of domestic trade. When including domestic trade, the case $i = j$ denotes domestic trade flows and $i \neq j$ denotes international trade flows. Moreover, in this case, the literature \citep{BergstrandLarchYotov:15} suggests to include an additional term in Equation~\ref{eq: bilateral trade costs} such as $\zeta_t b_{ij}$, identifying the ease of trading internationally versus trading domestically over time. This is a standard approach for disentangling the effect of trade globalization from the effect of other trade policy or monetary standard variables.} 
The term $Y_i \equiv \sum_j X_{ij}$ represents production in country $i$, while $E_j \equiv \sum_i X_{ij}$ represents expenditure in country $j$. 

As demonstrated by \citet{AndersonVanWincoop:03}, structural gravity models also satisfy two additional conditions:

\begin{equation} \label{eq: outward MR}
    \Omega_{i}^{-\theta} = \sum_{j} (\frac{\tau_{ij}}{\Pi_{j}})^{-\theta} \frac{E_{j}}{Y}
\end{equation}

and 

\begin{equation} \label{eq: inward MR}
    \Pi_{j}^{-\theta} = \sum_{i} (\frac{\tau_{ij}}{\Omega_{i}})^{-\theta} \frac{Y_{i}}{Y}
\end{equation}

The term $\Omega_{i}$ represents outward multilateral resistance and is specific to exporting country $i$, capturing its access to potential export markets. Conversely, $\Pi_j$ denotes inward multilateral resistance, reflecting the degree of competition that trade flows from any origin face in destination country $j$. Higher values of either term are associated with lower bilateral trade flows, which is why they are referred to as ``multilateral resistance terms''. The remaining component, $\tau_{ij}$, captures all pair-specific trade costs.

Exploiting the multiplicative structure of gravity models, and extending it to a panel setting with an additive error term, it is possible to reformulate Equation~\ref{eq: gravity theory} in a log-linearized form:

\begin{equation} \label{eq: loglinear}
X_{ijt} = \exp \left( \ln Y_{it} + \ln E_{jt} - \ln Y_{t} -\theta \ln \tau_{ijt} + \theta \ln \Omega_{it} + \theta  \ln \Pi_{jt} + \varepsilon_{ijt} \right).
\end{equation}

In Equation~\ref{eq: loglinear}, $\ln \tau_{ijt}$ is the only term that varies jointly by exporter and importer. All other terms depend solely on either the exporter or the importer. As a result, they can be collapsed into exporter-time and importer-time dummy variables, and absorbed in the estimation procedures by the corresponding fixed effects. These are the standard way of controlling for ``multilateral trade resistances,'' as defined by \citet{AndersonVanWincoop:03}. They also absorb all variables that vary at the exporter-time and importer-time level, such as GDP, GDP per capita, a country openness to trade, etc.

As the true vector of trade costs is not available to researchers, in gravity models the term $\ln \tau_{ijt}$ is specified using observable proxies. In this case, trade costs $\ln \tau_{ijt}$ are defined as follows:

\begin{equation} \label{eq: bilateral trade costs}
    -\theta \ln \tau_{ijt} = \beta_{LMU} {LMU}_{ijt} + \beta_{GS} {GS}_{ijt} +\gamma'{Other MS}_{ijt} + \beta_{TA} {TA}_{ijt} + \psi_{ij}.
\end{equation}

The term ${LMU}_{ijt}$ is my main (dummy) variable of interest, and identifies country-pairs pertaining to the LMU at time t. In a similar fashion, the term ${GS}_{ijt}$ controls for pairs where countries are both on gold standard. The vector of controls ${Other MS}_{ijt}$ identifies country pairs sharing the same monetary standard (other than gold or the LMU). Inspired by \citet{LopezCordovaMeissner:03} and \citet{MitchenerShizumeWeidenmier:10}, this corresponds to a set of dummy variables which are equal to one if both countries are on silver, on a bimetallic standard other than the LMU, or on paper, with the latter included for consistency with the monetary-regime classification used in this literature.\footnote{Since paper does not constitute a common monetary standard in the same sense as the metallic regimes, I also test the robustness of the results to excluding this category.}. The term ${TA}_{ijt}$ is a dummy variable that identifies trade agreements. As recently demonstrated by \citet{Timini:23}, trade agreements played a significant role in shaping trade during the 1860s and 1870s. It is therefore important to consider them as a potential confounding factor. The term $\psi_{ij}$ corresponds to directional country pair fixed effects, and mirrors the \citet{BaierBergstrand:07} approach to control for endogeneity in gravity models. In this case, these fixed also absorb time invariant bilateral trade costs.

In this first approximation, I will identify the ${LMU}_{ijt}$ effects in different periods, identified using the historical narrative detailed in the "Historical context" section, and following \citet{Timini:18}, by adapting the sample length. I will therefore exploit three different samples, ending at 1873, 1885, and 1913 respectively.

Despite being grounded on the historical narrative, in fact, the exact dates of these cuts is somewhat exogenously imposed by the researcher. Therefore, this can let the reader with doubts on whether the LMU trade effects really change around those years. To further address this issue, I will allow the coefficient of the LMU variable to vary over time in another specification. Furthermore, such specification will also serve to verify whether there are identifiable pre-trends on trade between LMU members, by backtracking the LMU variable to t-3. The time span is mostly dictated by the limited pre-LMU window available in the dataset (five years), which prevents a more extensive examination of pre-trends.\footnote{Including a longer set of pre-treatment leads would generate collinearity with the high-dimensional fixed effects, making it impossible to separately identify all pre-trend coefficients.} Given the short time period, pre-trends resulting from this exercise may also capture some anticipatory behavior, as negotiations and signatures usually precede the entry into force of an agreement.\footnote{The gravity literature is not unanimous on the number of years that should be checked before a trade agreement or a currency union enters into force, but it suggests that often some effects can be detected before formal implementation due to anticipation, rather than endogeneity. Recent contributions by \citet{EggerLarchYotov:22} and \citet{NagengastYotov:25} support this view, suggesting that the effects of trade agreements may begin up to three years prior to their official entry into force. In the specific case of the LMU, anticipation is historically plausible even if contemporaries did not necessarily foresee the exact institutional form of the 1865 convention. As \citet{Willis:01} shows, by the mid-1860s existing monetary difficulties had made concerted action appear increasingly as the natural solution. The idea itself was not new, as earlier experiences such as the German monetary convention of 1857 had already provided a precedent for international monetary coordination. Against this background, proposals for a union between Belgium, Italy, France, and Switzerland to regulate coinage jointly were being advanced from several months, before Belgium initiated formal discussions in early 1865. The 1864 coefficient may therefore be interpreted as consistent with expectations of some coordinated monetary response, without necessarily implying that agents anticipated the precise design eventually adopted by the LMU.}

Given sample length, reference years are, in this case, 1860 and 1861. Given the focus on a single currency union, the LMU, this is similar to running an event study that spans from t-3 to t+48.
In this sense, this corresponds to specify the vector of trade costs in the following form:

\begin{equation} \label{eq: bilateral trade costs_2}
    -\theta \ln \tau_{ijt} = \sum_{t=1862}^{1913} \beta_{t} {LMU}_{ijt} \cdot \mathbb{I}_{\{t = T\}} + \beta_{GS} {GS}_{ijt} +\gamma'{Other MS}_{ijt} + \beta_{TA} {TA}_{ijt} + \psi_{ij}.
\end{equation}

where $\cdot \mathbb{I}_{\{ \textit{cond} \}}$ denotes an indicator function that takes the value one if the condition $\textit{cond}$ is satisfied and zero otherwise.
In this case, the LMU variable starts three years earlier for each member, so to be able to identify the effects during the 1862-1864 period.

All estimations are performed following standard practice: I use Poisson pseudo maximum likelihood (PPML), as originally proposed by \citet{SantosSilvaTenreyro:06}. This method provides consistent parameter estimates and trade cost elasticities in the presence of zero trade flows and heteroskedasticity. As is standard practice, I use nominal trade data, following the recommendation of \citet{BaldwinTalgioni:07}, who argue that importer-year and exporter-year fixed effects adequately control for cross-country inflation differentials. Deflating the data using CPI would introduce unnecessary noise—an issue likely to be even more pronounced when working with 19\textsuperscript{th}-century trade records denominated in pounds, which would require applying the (reconstructed) Great Britain CPI deflator.

\subsection{Data}
In my main specification, I use data on gross international trade flows from version~4 of the TRADHIST database by \citet{FoquinHugot:16}. In some robustness checks, I also use the well-known historical trade database, RICardo \citep{DedingerGirard:17}, to verify the robustness of the results. TRADHIST database compiles historical bilateral trade flows of goods taken from various sources, including both primary sources and other trade databases, such as RICardo itself. Trade flows are nominal and are expressed in British pound sterling,\footnote{TRADHIST also allows researchers to compute domestic trade flows directly. Although domestic trade flows are not readily available from historical statistics, they can be constructed using TRADHIST as the difference between nominal Gross Domestic Product (GDP) and nominal total exports. Ideally, I would rely on gross total output statistics. However, these data do not exist for a large enough number of countries and years during our period of analysis. Therefore, GDP-based calculations are the best possible viable alternative. Importantly, a recent study by \citet{CamposTiminiVidal:21} shows that in well specified gravity models, exporter-time, importer-time, and pair fixed effects make the discrepancy between GDP (a measure of value added) and output (a gross measure) relatively unimportant in estimating many bilateral trade costs proxies, such as trade agreements or currency unions.} which provides a common nominal unit of account across countries and years.\footnote{Expressing trade flows in pounds sterling provides a common nominal unit of account. Any exchange-rate conversion factor applied consistently to a country’s reported trade flows in a given year is absorbed by the relevant exporter-year or importer-year fixed effect. The choice of sterling as the common numeraire therefore has no substantive effect on the estimated LMU coefficient.}

I follow the literature and identify gold standard membership using \citet{Officer:20}. Silver, bimetallic, and paper pairs (other than LMU countries) are identified using information on country currency standards contained in \citet{Meissner:24} and \citet{Meissner:01}. As in \citet{Flandreau:00} and \citet{Timini:18}, LMU member states include: Belgium, France, Switzerland, and Italy from 1865, with Greece joining in 1868.

The sample includes 37 countries and territories (see Table~\ref{tab:Countries_in_the_sample} in the Appendix for more details), and covers the period from 1860 to 1913.

\section{Results}
\subsection{Main results}
\label{sec:mainresults}

\begin{figure}[htbp]
\centering \subcaptionbox{Without controlling for other monetary standards}{\includegraphics[width=0.8\textwidth]{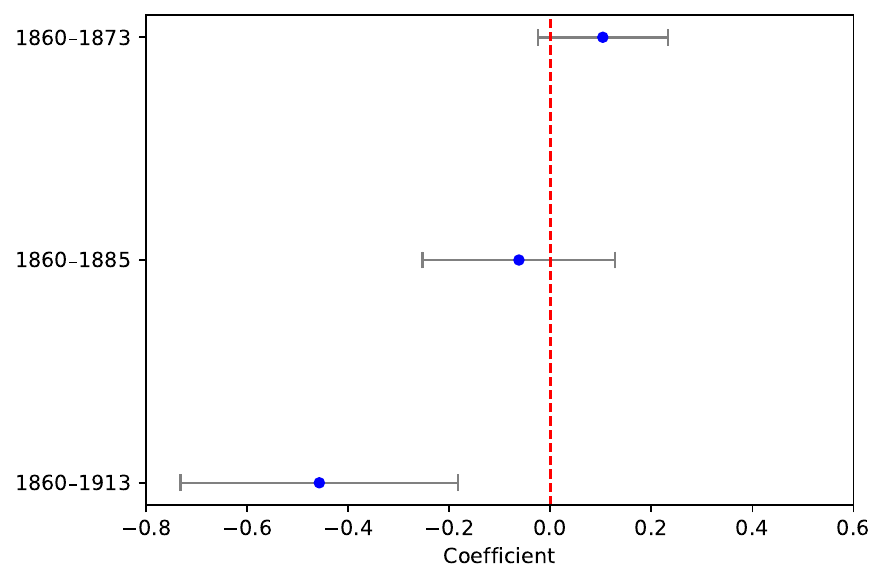}}
\\
\subcaptionbox{Controlling for other monetary standards}
{\includegraphics[width=0.8\textwidth]{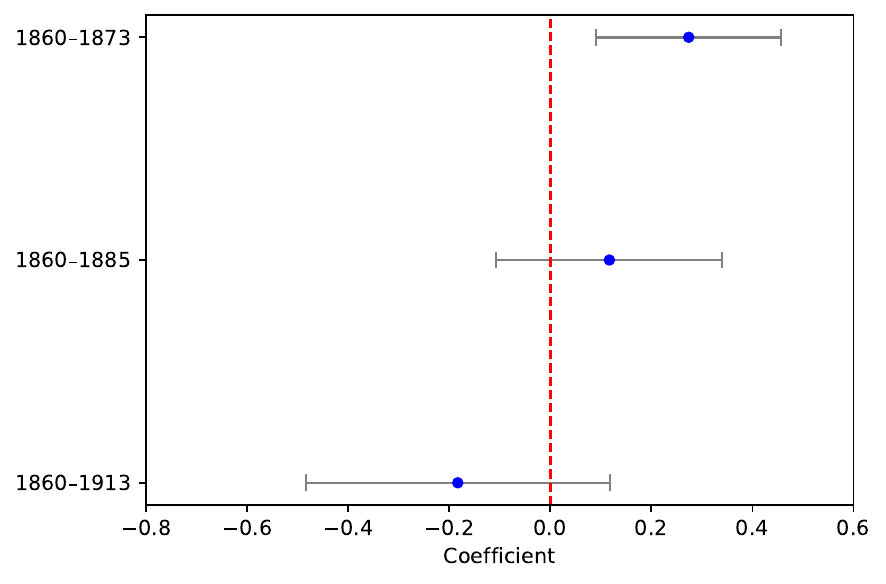}}
\caption{LMU trade effects with different control groups}
\label{fig: main_results}
\vspace{5mm}
\begin{minipage}[c]{0.8\textwidth}
\footnotesize
\textbf{Note}: The figures show point estimates (blue dots) and 90\% confidence intervals (grey lines). Standard errors are clustered by directional country pair. Estimations use data from the TRADHIST database. Both panels are based on regressions that specify the vector of trade costs based on Equation~\ref{eq: bilateral trade costs}. The top panel (``Without controlling for other monetary standards'') shows results from estimations that excludes the term $\bm{\gamma}' \bm{Other MS}_{ijt}$. The bottom panel (``Controlling for other monetary standards'') shows results from estimations that includes the term $\bm{\gamma}' \bm{Other MS}_{ijt}$.
\end{minipage}
\end{figure}

Figure~\ref{fig: main_results} reports the estimated effects of the LMU on trade. Panel (a) presents estimates from the baseline specification without controlling for other common monetary standards, in line with the approach taken in the previous literature \citep{Flandreau:00, Timini:18}. Results are reported for three samples: 1860--1873, 1860--1885, and 1860--1913. Relative to both country pairs without a shared monetary standard and those sharing a monetary standard other than gold (e.g., silver or bimetallic standards outside the LMU), the LMU has only a small and statistically insignificant effect on trade.
However, when these other standards are explicitly accounted for (see Panel (b)), the estimated LMU effect changes substantially. This is particularly relevant for the early years of the LMU (until 1873), a period where many different currency regimes coexisted, and it is consistent with the approach generally used in studies of the gold standard and twentieth-century currency unions. Failing to account for these regimes can bias the LMU coefficient by placing pairs already benefiting from monetary coordination in the control group.

The estimates suggest that the LMU increased trade between its members by approximately 30\% compared to country pairs that did not share a monetary standard. This control group is comparable to those employed in most studies of the gold standard period or 20\textsuperscript{th}-century currency unions. It allows for a cleaner comparison of the LMU effect relative to country pairs that did not share the same monetary standard. These were precisely the pairs that remained exposed to the monetary frictions the LMU was supposed to mitigate, such as uncertainty regarding coinage standards, cross-border acceptability, and the risk of unilateral changes in monetary conduct.\footnote{Exact coefficients and standard errors for the LMU dummy and the control variables are reported in Appendix Table \ref{tab:lmu_trade}. The specification is designed to identify the LMU coefficient; the remaining monetary standard dummies are included to account for alternative monetary arrangements and should be interpreted as conditional associations rather than as separate causal estimates. Reassuringly, their signs and magnitudes are broadly consistent with the previous literature. I discuss the interpretation of these controls below Table \ref{tab:lmu_trade} and Table \ref{tab:rob_lmu_trade_1860_1873} in the Appendix.}

Figure~\ref{fig:lmu_over_time} shows the results of the estimation procedure using the trade costs specification reported in Equation~\ref{eq: bilateral trade costs_2}. In this way, I go beyond the partition of periods ``exogenously'' imposed by the researcher, and I am able to portray the evolution of the LMU trade effect over time. There are at least four important points to note. First, between its creation in 1865 and the suspension of free silver coinage in late 1873 and early 1874, the LMU had positive and significant effects on trade among its members. Second, these effects started to fade at least from 1873 onward. This decline coincides with the restrictions on free silver coinage introduced in 1873–1874, which weakened the bimetallic mechanism on which the LMU had been based. By around 1878, the year in which the
minting of new silver coins was finally stopped throughout the LMU, they had approached zero.
Third, from a trade perspective, the LMU was already dead well before the time of the 1885 reform---what \citet{Willis:01} suggested was the \textit{de facto} ``abrogation of the Latin Union''.
Fourth, these findings also suggest that the specification implemented in the paper plausibly captures endogeneity well. The coefficients reported for 1862-1864 (three year window before the entry into force of the LMU) show a small and non statistically significant coefficient.\footnote{Although the coefficient for 1864 is twice as large as those for 1862 and 1863, this pattern is common in studies of trade agreements and currency unions, where effects often precede formal implementation due to anticipatory behavior, as negotiations and signatures typically occur one to three years before entry into force. Recent contributions by \citet{EggerLarchYotov:22} and \citet{NagengastYotov:25} support this view, suggesting that the effects of trade agreements may begin up to three years prior to their official entry into force. While the limited pre-LMU window in the dataset constrains further backward extension—--at least without unbalancing the dataset as observations for key countries, including LMU members, such as Italy, do not exist before the 1860s--—the already small and non-statistically significant coefficients for 1862 and 1863 are reassuring.}

A possible concern is that the LMU coincided with other forces that may have strengthened trade ties among several European economies during the 1860s and the early 1870s. These include the broader wave of trade liberalization following the Cobden---Chevalier Treaty, Italy's post-unification integration into European markets, and the political and diplomatic centrality of France within the franc area. The empirical specification accounts for several of these concerns. Exporter-year and importer-year fixed effects absorb all country-specific developments that vary over time, including changes in output, expenditure, industrialization, openness, domestic institutions, monetary or fiscal conditions, and country-wide shocks affecting each country’s capacity to export or import. Among these, they also absorb country-wide trade policy changes, such as unilateral or multilateral liberalization episodes, MFN-type tariff reductions, or other policy changes affecting all trading partners of a given country in a given year. The specification then explicitly controls for trade agreements, which is particularly important given the broader liberalization wave of the 1860s, as these capture bilateral preferential liberalization not absorbed by exporter-year and importer-year fixed effects. Additionally, directional country-pair fixed effects absorb all time-invariant bilateral determinants of trade, such as distance, common borders, language, and historical ties. In this sense, the estimated LMU effect is not identified from pre-existing similarities between member countries---such as geographical proximity, common language, institutional similarity, or already strong trade integration---as these bilateral characteristics are absorbed by the pair fixed effects. While I further assess the robustness of the results to additional political controls (such as military alliances), alternative samples, and methodological choices, these checks cannot, however, fully disentangle the LMU effects through the monetary channel from LMU-wide political, financial, or diplomatic forces. This is a common issue in the literature on both historical and contemporary currency unions and monetary agreements, where participation in a monetary arrangement may capture not only monetary frictions, but also the broader institutional and political environment in which such arrangements are embedded.

The LMU coefficient should therefore be interpreted as the average trade effect associated with membership in this institutional framework, rather than as a direct estimate of any single mechanism. Common coinage rules may have lowered transaction costs by making coins more easily interchangeable and reducing uncertainty about their future acceptance. These effects could have persisted even with imperfect compliance, as long as deviations were regarded as manageable and confidence in bimetallism remained intact. Broader institutional and diplomatic factors may also have played a role. France occupied a central position within the franc area, not only because the Union was built around the franc-based coinage system, but also because French authorities viewed monetary coordination as part of a wider diplomatic and political project. As emphasized by \citet{Einaudi:01}, the LMU was seen in France as a possible step toward broader European monetary unification, and such debate involved governments, economists, bankers, and commercial interests across countries. 
The concentration of the estimated effect in the Union’s early years is consistent with these channels weakening as internal strains and the broader crisis of bimetallism reduced confidence in the arrangement.

\begin{figure}[htbp]
    \begin{center}
    \includegraphics[width=0.95\textwidth]{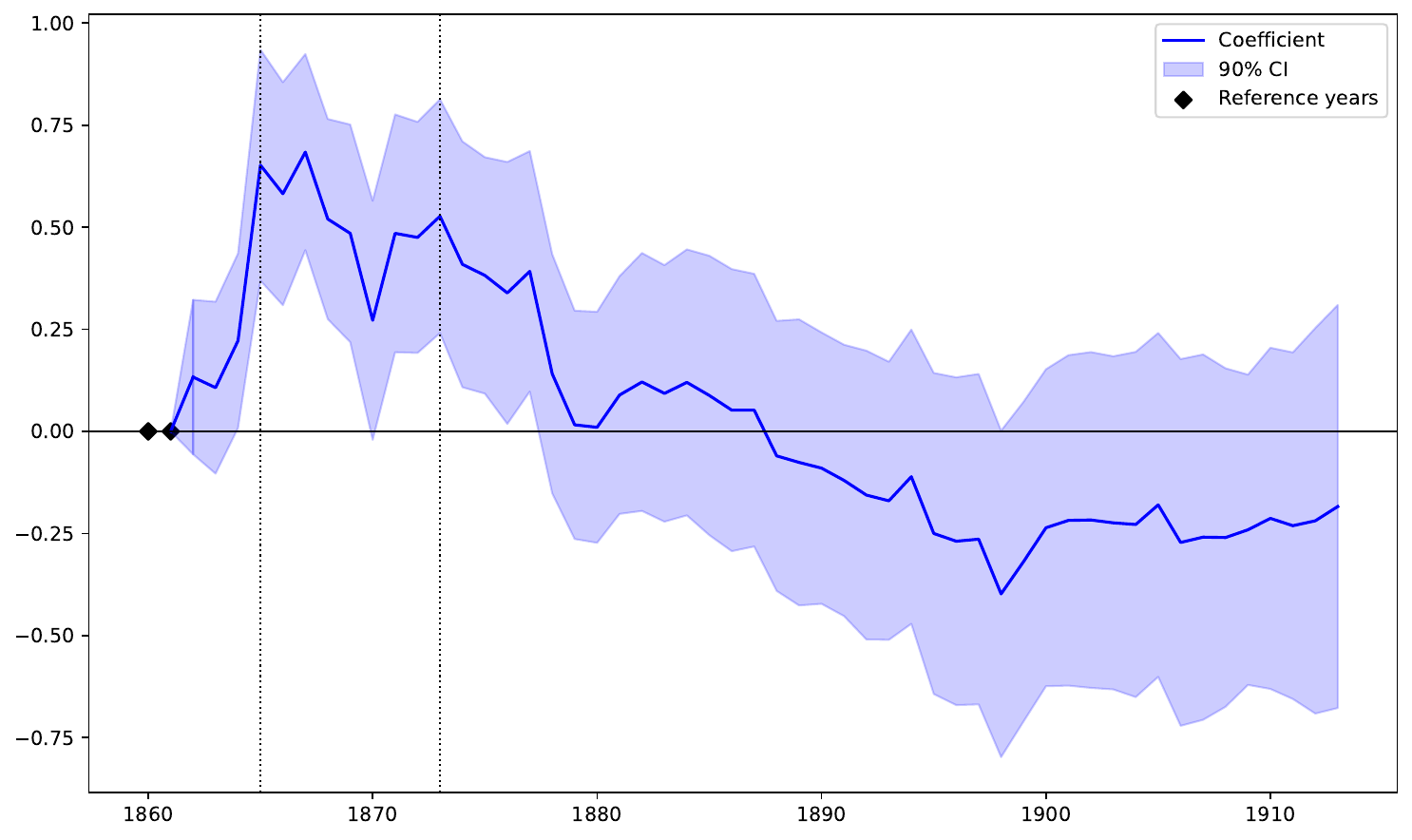}
    \caption{The Latin Monetary Union trade effects}
    \label{fig:lmu_over_time}
    \vspace{3mm}
        
    \begin{minipage}{0.9\textwidth}
    \footnotesize
    \textbf{Note}: The figure shows the estimated coefficient of the LMU over time ($\hat{\beta}_t$) and 90\% confidence intervals. Standard errors are clustered by directional country pairs. Estimations use data from the TRADHIST database. The estimation uses the trade costs specification reported in Equation~\ref{eq: bilateral trade costs_2}.  
    \end{minipage}
    \end{center}
\end{figure}

The fact that the estimated effects are concentrated in the years in which bimetallism still retained credibility, and fade after the suspension of free silver coinage, links the trade results directly to the monetary chronology outlined by \citet{FlandreauOosterlinck:12} and \citet{Flandreau:96}. While they emphasize the impact of the suspension of free silver coinage in 1873-74 on the viability of bimetallism, this paper suggests that these events also had significant consequences for trade among LMU members.\footnote{As discussed, these findings are consistent with the monetary chronology documented by \citet{FlandreauOosterlinck:12} and \citet{Flandreau:96}. However, they differ from those reported in \citet{Vicquery:21}. This difference likely reflects variation in scope rather than a direct inconsistency between the results. In particular, \citet{Vicquery:21} focuses on trade between selected Italian pre-unitary states and their LMU partners around the time of Italian unification, using a dataset limited to maritime trade and ending in 1869. As a result, the analysis covers only a short period following the creation of the LMU and only a portion of intra-LMU trade flows.} 

These findings have two broader implications. First, the results qualify the existing evidence on the LMU. Earlier studies found negligible or modest trade effects overall, with the larger estimates confined to specific bilateral flows. Once the control group distinguishes pairs sharing other monetary regimes from those without a common arrangement, the estimated average LMU effect is positive during its early years. Second, the results have a methodological implication for studies of 19\textsuperscript{th} century trade. When several monetary regimes coexist, it is important to control for them even when the analysis focuses on other international policies. Otherwise, the control group may combine country pairs subject to different forms of monetary coordination, making the effect of the policy under study harder to identify.

\subsection{Robustness tests}
Results are robust across a range of specifications, including the consideration of additional potential confounders, and the use of alternative methodological choices, different samples, or trade data sources. Importantly, given that the study concerns a currency union designed to facilitate the unrestricted circulation of coins across national borders, I also explore here how the observed results are not driven by underlying specie flows.

More in detail, in the spirit of \citet{GowaHicks:13}, \citet{GowaHicks:17}, and \citet{KarlssonHedberg:21}, I additionally control for military alliances between two countries, and for whether two countries are at war with each other.\footnote{More precisely, data on military alliances are based on \citet{Gibler:09} and are codified as a dummy variable equal to one if two countries (the exporter and the importer) have a defense pact in force at time t, and zero otherwise. In the database, a defense pact is defined as ``the highest level of military commitment, requiring alliance members to come to each other’s aid militarily if attacked by a third party''. Interstate dyadic war data are based on \citet{MaozJohnsonKaplanOgunkoyaShreve:19}, and are codified as a dummy variable equal to one if two countries (the exporter and the importer) are at war with each other at time t, and zero otherwise.} 

I then experiment with different methodological choices, such as including both international and domestic trade in the left-hand side variable \citep{Yotov:22},\footnote{When domestic trade flows are included, the gravity literature recommends adding time-varying indicators for international trade \citep{BergstrandLarchYotov:15, LarchShikerYotov:25}. These indicators vary by year and take value one for international flows and zero for domestic flows. Their purpose is to absorb changes over time in the relative importance of international trade, such as those generated by declining transport and communication costs, trade liberalization, or broader globalization forces. Including them ensures that common changes in international integration are not attributed to the bilateral variables of interest, such as the LMU in this case.} treating the LMU and the gold standard as mutually exclusive monetary arrangements,\footnote{In the baseline specification, LMU country-pairs keep the LMU dummy equal to one as long as the Union formally existed. However, from 1878 onward, LMU members are also coded as belonging to the gold standard, since free silver coinage had been halted and the Union had effectively moved to a ``limping gold standard''. The baseline specification therefore allows the LMU and gold standard dummies to overlap in the late period. As a robustness check, I impose a mutually exclusive coding by setting the gold standard dummy equal to zero whenever the LMU dummy is equal to one.} correcting for possible biases in the estimating procedure \citep{WeidnerZylkin:21}, and testing alternative clustering strategies as suggested by \citet{EggerTarlea:15}. I also consider an alternative coding of monetary standards motivated by \citet{Flandreau:04}. His analysis emphasizes that exchange-rate volatility across metallic regimes was relatively low before the 1870s. I therefore add a dummy for pairs in which both countries were on metallic standards---gold, silver, bimetallic, the LMU, or another bimetallic standard---but did not share the same specific regime. I also remove paper--paper pairs from the vector of other common monetary standards, so that the omitted category consists of pairs in which at least one country was on paper. This keeps the control group confined to pairs outside the relatively stable environment of metallic regimes.

I also test whether the results are sensitive to the sample or trade data used. 
First, I extend the sample back to 1855, which provides a longer pre-LMU window. However, this exercise comes at a cost, since the earlier years are more unbalanced and necessarily exclude Italy, one LMU member, as unified Italian trade data are not available before the 1860s.
Second, I address the specific role of Italy. Italy suspended convertibility in 1866, shortly after joining the LMU, making it a potentially important qualification to the credibility channel emphasized above. I therefore re-estimate the baseline specification excluding Italy from the sample. The estimated LMU coefficient remains very similar in magnitude to the baseline estimate. This suggests that the early positive trade effect is not driven by Italian trade flows or by Italy-specific monetary developments after 1866. Rather, the result appears to capture a broader LMU-wide effect during the period in which the bimetallic framework still retained credibility.
Third, I adapt the countries included in the sample according to the previous literature on trade during the first globalizaton \citep{LopezCordovaMeissner:03, Timini:18, Timini:23}. Additionally, I also consider other sample restrictions. These include limiting the sample to European countries only; excluding Germany; and focusing on European countries and the United States. Finally, I use RICardo \citep{DedingerGirard:17} as an alternative database for sourcing information on bilateral trade flows.

\begin{figure}[htbp]
    \begin{center}
    \includegraphics[width=0.95\textwidth]{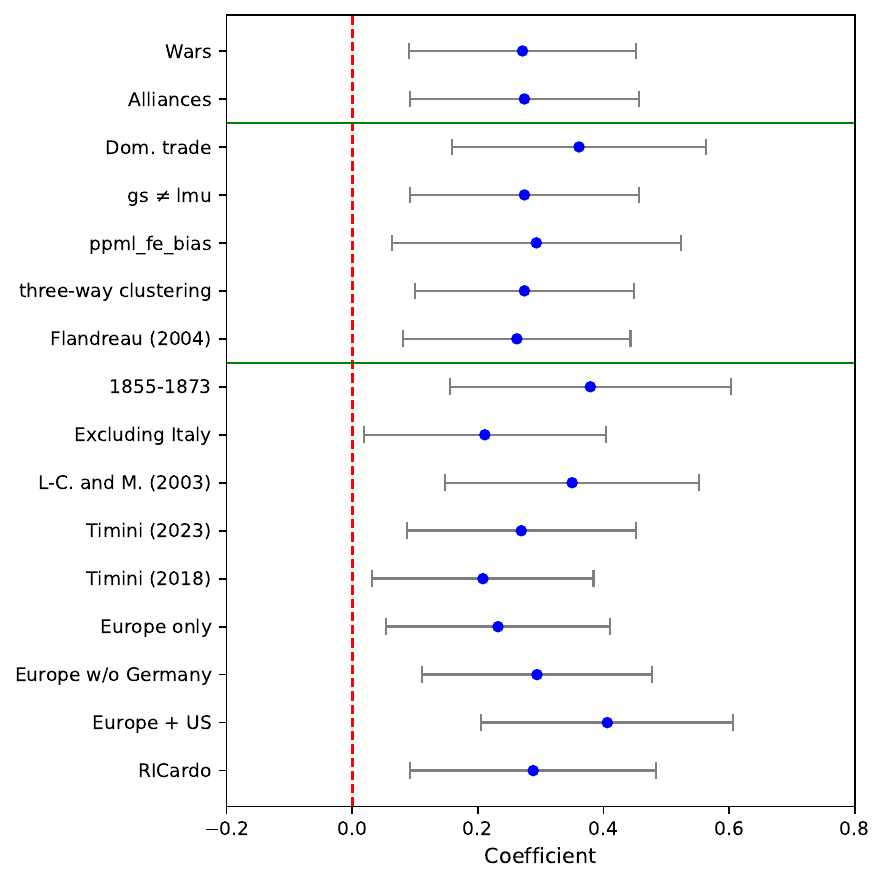}
    \caption{Robustness tests}
    \label{fig:lmu_robustness}
    \vspace{3mm}
        
    \begin{minipage}{0.9\textwidth}
    \footnotesize
    \textbf{Note}: The figures show point estimates (blue dots) and 90\% confidence intervals (grey lines). Standard errors are clustered by directional country pair. Estimations use data from the TRADHIST database. The panel is based on regressions that specify the vector of trade costs based on Equation~\ref{eq: bilateral trade costs}. ``L-C. and M. (2003)'' stands for \citet{LopezCordovaMeissner:03}''. See text for more details.
    \end{minipage}
    \end{center}
\end{figure}

Figure~\ref{fig:lmu_robustness} reports robustness test results for the LMU effects during the 1865-1873 period. The estimated coefficient ranges from 0.21 to 0.41, indicating a trade increase of between +23\% and +50\% depending on the specification used.\footnote{Exact coefficients and standard errors for the LMU dummy and the control variables in all robustness exercises shown in the Figure, are reported in Appendix Table \ref{tab:rob_lmu_trade_1860_1873}. I discuss the interpretation of these controls below Table \ref{tab:lmu_trade} and Table \ref{tab:rob_lmu_trade_1860_1873}.} 

\begin{figure}[htbp]
    \begin{center}
    \includegraphics[width=0.95\textwidth]{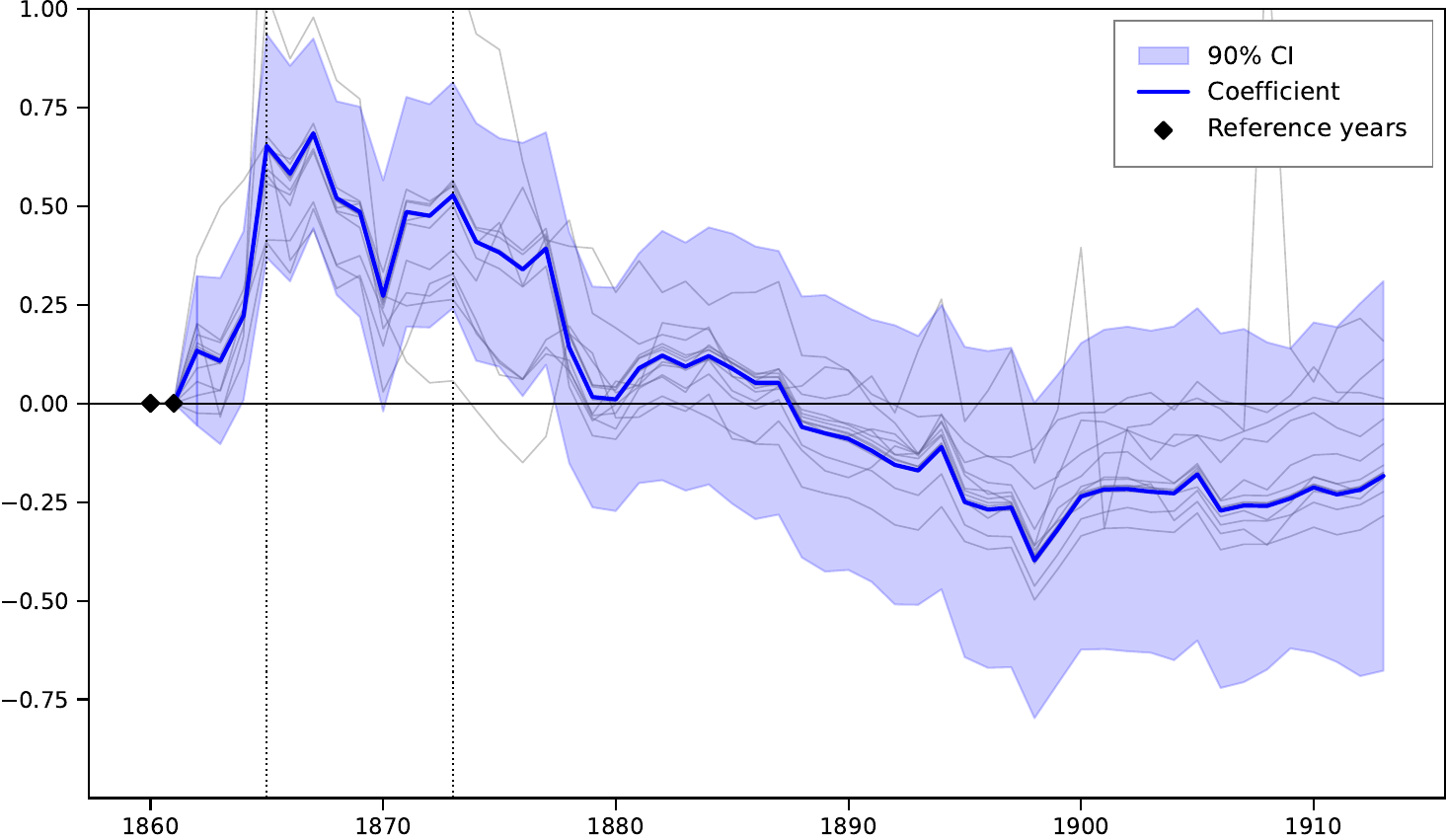}
    \caption{Robustness tests (LMU effect over time)}
    \label{fig:lmu_robustness_overtime}
    \vspace{3mm}
        
    \begin{minipage}{0.9\textwidth}
    \footnotesize
    \textbf{Note}: The figures show point estimates and 90\% confidence intervals for the main specification. Standard errors are clustered by country pair. Light gray lines report coefficient estimates of the robustness tests described in the text, based on regressions that specify the vector of trade costs based on Equation~\ref{eq: bilateral trade costs_2}. See text for more details. Standard errors are clustered by directional country pairs. 
    \end{minipage}
    \end{center}
\end{figure}

Figure~\ref{fig:lmu_robustness_overtime} reports robustness test results for regressions that allow the LMU coefficient to vary over time. The Figure displays the coefficients and confidence intervals from the main regressions, and then report coefficients from all robustness tests.\footnote{In this case, the \citet{WeidnerZylkin:21} methodology is not implemented as it does not allow for multiple coefficients of interest in the same regression. Also, results from alternative clustering strategies \citep{EggerTarlea:15} are not reported as---by construction---coefficients are identical to the main specification, i.e. only standard errors change.} Quite nicely, practically all coefficients from the robustness tests fall within the confidence intervals of the coefficient of the main regressions. One exception is the RICardo-based regression, where the distribution of coefficients over time is more noisy. However, point estimates tend to be larger (and not smaller), falling above the confidence intervals (for the coefficient estimated in the main regression) for the period 1865-1873.\footnote{However, it is possible that this more noisy distribution and the larger coefficients when using the RICardo database are dictated by a lower number of observations in certain years and pairs.}  

Finally, while historical (as well as contemporary) trade statistics record merchandise flows, these often include specie movements. Nevertheless, the interpretation of the role of specie flows remains underexplored. The ambiguity lies in whether specie was actively traded as a good or passively moved to settle imbalances in the current or capital accounts. While this distinction is hopefully of second order importance in most cases---especially when studying trade policy issues, or other forms of trade integration---it could be a more prominent issue for the study of the LMU, as it was explicitly designed to facilitate the unrestricted circulation of coins across national borders.

Ideally, to make sure that my main results are not driven by specie flows, I would like to have access to bilateral trade data disaggregated by product for a large set of countries. This would allow me to separate merchandise trade from specie flows directly.\footnote{Technically, as most historical trade databases have been compiled using SITC Rev. 2 classification, this implies to exclude items 96 and 97 at the SITC 2-digit level.} 
There are only a few countries for which such database is available, and, to the best of my knowledge, Italy is the only one for which such database is granted public access, covering the period 1862-1939.\footnote{\citet{JonesObstfeld:00} construct aggregate country-level series for gold flows and non-gold current accounts for a number of countries. Although useful for the analysis of aggregate external accounts, these data do not report specie and merchandise trade by trading partner and therefore cannot be used to remove specie flows from bilateral trade observations.}

Therefore, as a second-best strategy to support the statement that my main results are not driven by underlying specie flows, I use Italian product-level data \citep{FedericoNatoliTattaraVasta:12}, and check that specie flows are not an important component of trade flows among LMU members for the period 1865-1873, i.e. when I find positive and significant results of the LMU on trade. Reassuringly, specie flows correspond to less than 0.8\% of total bilateral trade flows between Italy and other LMU members.
This is further confirmed by a set of regressions explicitly identifying trade data that include specie flows.\footnote{This consist in adding to my baseline specification, for the period 1865-1873, a dummy equal to one when trade data sources explicitly acknowledge the inclusion of species and bullions in trade data, on the basis of the information provided by \citet{DedingerGirard:17} in the RICardo database. Given that the source is linked to the RICardo database, I run two different regressions where the trade flow variable is respectively based on the information contained in TRADHIST and RICardo. The estimated LMU coefficient corresponds to 0.255 and 0.288 respectively, and is always significant at the 5\% level.}

\subsection{Extension: Core - Periphery (or the LMU's heterogeneous trade effects)}
\label{sec:core_per}

Previous contributions analyzed the possible existence of heterogeneity in the LMU trade effects across its members using econometric methods. \citet{Timini:18} focuses on differences in trade flows between France and other members, and those involving other members only. He does so by splitting the LMU dummy into separate dummies. \citet{Vicquery:21}, instead, looks separately at the bilateral flows between each LMU members and some Italian pre-unitary states.
The heterogeneous effects of different trade agreements or currency unions have been assessed in literature using gravity models also for the post-WWII period \citep{BaierYotovZylkin:19, GlickRose:16}, though authors often acknowledge the limitations of such extensions.\footnote{If applying the econometric strategy suggested by \citet{Timini:18} to my data, however, results are similar to those of \citet{Timini:18} and available upon request.}
As noted by \citet{BaierYotovZylkin:19}, such disaggregation can dilute the identification strategy and compromise the robustness of the estimates: the more granular the estimate obtained---i.e. the fewer data points and countries involved in generating it---the wider the confidence bands of the coefficient, the higher the likelihood of incurring in an omitted variable bias or reverse causality.  

Here, I therefore evaluate whether the LMU trade effects were heterogeneous across countries exploiting the LMU-wide econometric estimates discussed in Section~\ref{sec:mainresults}, derived from Equation~\ref{eq: bilateral trade costs}, within a model-based approach, grounded in trade theory. There are two main reasons for preferring this approach. First, the model extends beyond partial equilibrium bilateral trade effects to assess the effects within a general equilibrium framework. By doing so, it also accounts for possible trade diversion effects, and therefore allows to calculate with more precision the total (and not only the bilateral) trade gains. Second, as it is based on LMU-wide estimates, it also minimizes the concerns noted in the previous paragraph.

One of the beauties of this class of general equilibrium models---based on the exact-hat algebra approach of \citet{DekleEatonKortum:08} and \citet{CaliendoParro:15}, and encompassing more general “universal gravity” formulations such as \citet{AllenArkolakisTakahashi:20}---is that it allows to capture heterogeneous trade effects across countries even when a uniform reduction in trade costs across all country pairs within a currency union (or a trade agreement) is used. Indeed, the ``universal gravity'' is a powerful framework for economists seeking to understand how trade flows respond to changes in trade costs. When used in counterfactual simulations—--such as changing trade barriers within the LMU—--the model shows how trade between countries adjusts based on the structure of existing trade relationships. Crucially, it does this in a transparent and data-driven way: the size of country-level effects depends on how much countries involved in trade costs reductions already trade with each other. If two countries are major trading partners, a reduction in trade costs will lead to a large increase in their total trade; if their trade is minimal to begin with, the effect will be modest. This intuitive mechanism makes the model especially appealing for historical analysis, as it allows researchers to isolate the impact of trade cost changes within a rigorous yet tractable framework, without needing to model preferences, technologies, or other more complex dynamics.\footnote{For the details of the model, see Appendix~\ref{sec:theory_appendix}.}

Therefore, building on this framework, it is possible to implement counterfactual simulations. The baseline consists of observed trade flows recorded in the database, i.e. the actual transactions that occurred between 1865 and 1873.\footnote{As the model is a static model, I use the average of the flows between those years.} The counterfactual calculates what trade would have taken place had Belgium, France, Greece, Italy, and Switzerland not joined the LMU. The difference between these two measures reveals the part of trade attributable to the LMU.

The simulation exercise requires only four key inputs: a complete and square bilateral trade matrix for a chosen baseline year, a specified change in trade costs, and two elasticity parameters---the trade elasticity and the supply elasticity. The former (trade elasticity) tells us how much trade between countries increases or decreases in response to a change in trade costs, while the latter (supply elasticity) reflects how producers adjust their output when export prices change.\footnote{I run the counterfactual simulation using the \texttt{ge\_gravity2} command \citep{CamposReggioTimini:25}. This command allows users to compute counterfactual trade flows in a large class of general equilibrium trade models in Stata.}

Therefore, for the bilateral trade matrix, I build upon the one employed in partial equilibrium estimations.\footnote{To obtain a complete and square bilateral trade matrix I first set to zero missing trade data for the first year of the database, 1860. Second, I interpolate (and extrapolate) existing trade data. Third, I take the average of the resulting trade flows between 1865 and 1873. This strategy is unlikely to distort the results of the general equilibrium trade model used. Indeed, in this class of trade models, what fundamentally drives outcomes are relative trade costs and trade shares, not the absolute levels of trade flows. The model can recover meaningful counterfactuals provided that the relative pattern of bilateral trade is preserved. Setting missing 1860 flows to zero and interpolating the data improves coverage, while averaging over 1865–-1873 reduces short-term noise and provides a more representative picture of trade relationships.}. The change in trade costs is derived by making use of the structural estimates obtained in partial equilibrium, under the standard assumption of symmetry.\footnote{See, for example, \citet{MayerVicardZignagoJavorcik:19}. This means that I treat the effects of joining and leaving the LMU as mirror images---the reduction in trade costs associated with LMU membership is assumed to be equal in magnitude and opposite in sign to the increase in trade costs following exit.} More precisely, this means that the change in bilateral trade costs in the counterfactual world without the LMU is given by

\begin{equation*}
    \hat{\tau}_{ijt} =
    \begin{cases}
	\exp \left(-\hat{\beta}_1/\theta\right), & \text{if $i$ and $j$ both identify LMU members,}\\
        1, & \text{otherwise.}
    \end{cases}
\end{equation*}

Trade and supply elasticities are calibrated based on values commonly used in the international trade literature. Specifically, I set the trade elasticity, $\theta$, to 5, similar to the value contained in the famous meta-analysis by \citet{HeadMayer:14}. The supply elasticity is set to 1.24.\footnote{However, since the text only discusses counterfactual results for trade, the choice of supply elasticity is less relevant, as this parameter is particularly important for welfare.} This value has been computed by \citet{CamposEstefania-FloresFurceriTimini:23} using statistics from \citet{HuoLevchenkoPandalai:23}, and positions my calibration between the benchmarks established by \citet{EatonKortum:02} and \citet{AlvarezLucas:07}.\footnote{Trade deficits are also an integral ingredient of the model to solve counterfactual simulations. As explained in \citet{HeadMayer:22}, there is no fully satisfying way to model trade deficits in a static model. A common assumption in the literature, which I also adopt in this paper, is that trade deficits are fully exogenous and therefore constant. Another possible assumption, often referred to as ``multiplicative'' assumption in the literature, is that trade deficits increase automatically with income (though not necessarily at the same rate). While these assumptions are theoretically different, changing them produces very similar results. Therefore, the choice is not particularly relevant in practical terms.}

\begin{figure}[htbp]
\centering \subcaptionbox{Levels, millions (£)}{\includegraphics[width=0.45\textwidth]{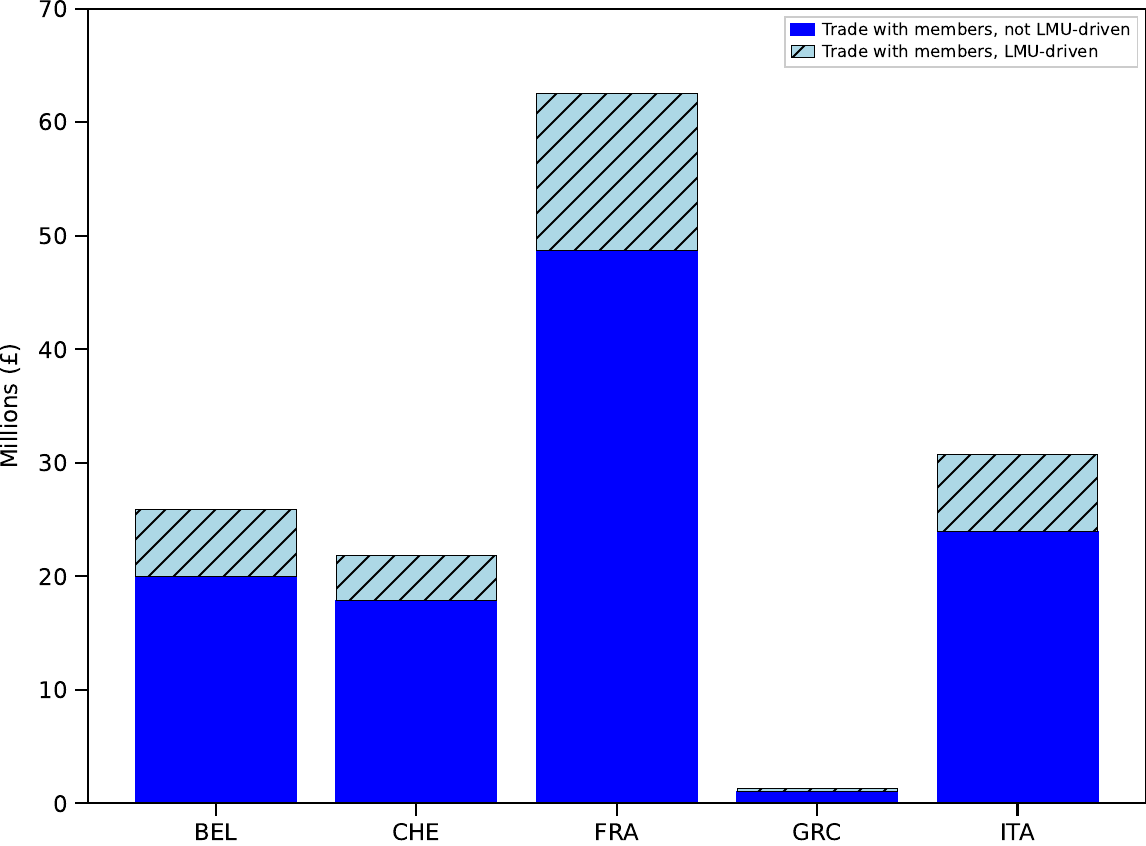}}
\subcaptionbox{\% of total trade}
{\includegraphics[width=0.45\textwidth]{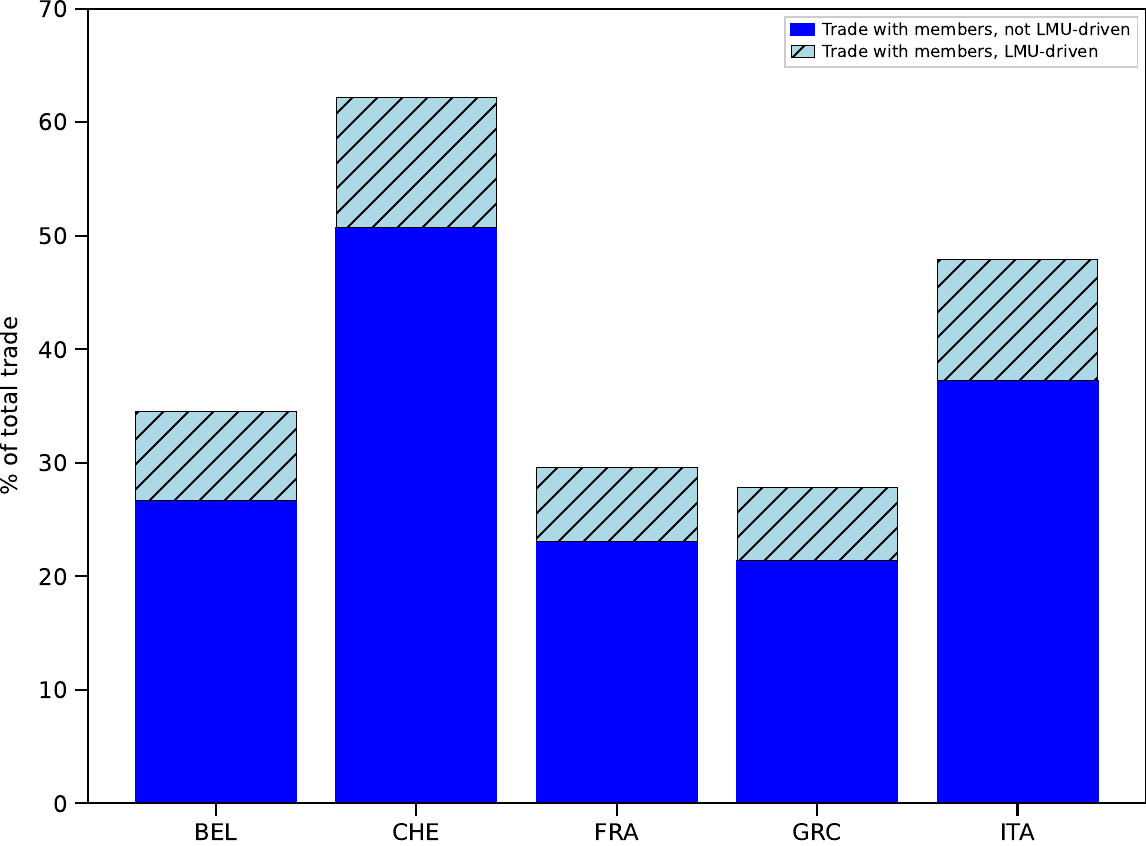}}
\caption{LMU-driven trade increase by country}
\label{fig: ge_results}
\vspace{5mm}
\begin{minipage}[c]{0.9\textwidth}
\footnotesize
\textbf{Note}: This figure reports model-implied trade attributable to the Latin Monetary Union (LMU), computed within a multi-country general equilibrium trade framework commonly referred to as ``universal gravity'' \citep[e.g.][]{AllenArkolakisTakahashi:20}. The baseline economy is constructed from observed bilateral trade flows in the database: the baseline trade matrix used in the simulation is the average of the completed bilateral flows over 1865--1873 (static model). The counterfactual economy holds all fundamentals fixed but assumes that Belgium, France, Greece, Italy, and Switzerland do not join the LMU. This is implemented as a uniform increase in bilateral iceberg trade costs for LMU pairs only. Under the standard symmetry assumption, counterfactual trade costs are set to $\hat{\tau}_{ij}=\exp(-\hat{\beta}_1/\theta)$ if both $i$ and $j$ are LMU members, and to 1 otherwise, where $\hat{\beta}_1$ is the LMU trade-cost estimate from Eq.~\ref{eq: bilateral trade costs} and $\theta$ is the trade elasticity. Equilibrium counterfactual trade flows are computed using the \texttt{ge\_gravity2} command \citep{CamposReggioTimini:25}, which implements a large class of general equilibrium gravity models via hat algebra. The trade elasticity is set to $\theta=5$ and the supply elasticity to 1.24; trade deficits are treated as exogenous and held constant across baseline and counterfactual. ``LMU-driven'' trade is defined as the difference between observed intra-LMU trade and the model-implied counterfactual without the LMU. Panel (a) reports this difference in levels, while panel (b) expresses it as a share of observed intra-LMU trade.
\end{minipage}
\end{figure}

The results of the general equilibrium model are shown in Figure~\ref{fig: ge_results}. The Figure shows the proportion of international trade among LMU members that the model attributes to the LMU (``LMU-driven''). The simulation shows that the LMU was an important positive factor for all its members. However, it also shows that its effects were uneven across countries. While France stands out in absolute terms (panel a), the relative trade gains (panel b)---expressed as percentage of each member’s total international trade with all partners, including non-LMU countries---of France (6.6\%) and Greece (6.5\%) were between a half and two-thirds the size of those of Switzerland (11.4\%) and Italy (10.7\%). Belgium stands somewhere in the middle (7.9\%). The cross-country variation therefore reflects both general equilibrium adjustments and differences in members’ exposure to intra-LMU trade.\footnote{The reported magnitudes are computed from counterfactual trade flows, and thus capture changes in trade values. They are expressed as shares of each member's total international trade, including trade with non-LMU countries. 
When evaluated over intra-LMU trade alone, the implied aggregate increase is only somewhat below the 30\% bilateral partial equilibrium estimate in Section~4.1. The small difference indicates limited general equilibrium adjustment, including trade diversion. In the single-sector CES model, expenditure shifts across country varieties and prices, output, and multilateral resistance adjust jointly. Given the observed trade shares and calibrated elasticities, the net effect on aggregate intra-LMU trade is modest.}

\clearpage

\section{Conclusions}
The LMU was a 19\textsuperscript{th} century agreement among several European countries to standardize their currencies through a bimetallic system based on fixed gold and silver content.
In this paper, I analyze its effects on trade using state-of-the-art gravity techniques and, crucially, accounting for the diversity of currency regimes during the early years of the LMU---therefore addressing a potential source of omitted variable bias.

The evidence presented in this paper suggests that the LMU had a meaningful, though short-lived, impact on trade among its members. In the initial phase following its establishment (1865–1873), when bimetallism remained a credible monetary arrangement, trade among its members increased by approximately 30\%. These effects then started rapidly fading, converging to zero by the end of the 1870s, when new silver coin minting was halted across the LMU. A model-based assessment suggests that, between 1865 and 1873, the LMU generated positive but uneven gains across its members. Switzerland and Italy experienced the largest relative gains, followed by Belgium, France, and Greece.

The paper also has a broader methodological implication. During the early LMU period, several monetary regimes coexisted, with some non-LMU pairs sharing gold, silver, bimetallic, or paper regimes. Treating all non-LMU pairs as a single control group therefore combines pairs with and without alternative forms of monetary coordination. Distinguishing among these regimes produces a control group more consistent with those used in studies of the gold standard and modern currency unions. The positive early LMU estimate helps bring the LMU evidence into line with the broader literature, which generally finds that common monetary arrangements promoted trade. More broadly, the results show that the composition of the control group can be important for estimating the trade effects of other international policies.

Similar methods could be applied in future research to examine the impact of other historically important exchange rate regimes or trade policy choices on trade.

\end{doublespace}
\clearpage

\bibliographystyle{ecta}
\bibliography{biblio}

\appendix
\clearpage
\counterwithin{figure}{section}
\counterwithin{table}{section}

\section{Summary statistics}

\begin{table}[ht]
   \centering
    \caption{Countries included in the sample: ISO codes} \label{tab:Countries_in_the_sample}
\begin{tabular}{lcccccc} \hline
ARG & AUS & AUT & BEL & BGR & BRA \\
\hline
CAN & CHE & CHL & CHN & COL & CUB \\
\hline
DEU & DNK & EGY & ESP & FIN & FRA \\
\hline
GBR & GRC & IDN & IND & ITA & JPN \\
\hline
KOR & MEX & NLD & NOR & NZL & PHL \\
\hline
PRT & RUS & SWE & TWN & URY & USA \\
\hline
ZAF &     &     &     &     &     \\
\hline
\end{tabular}
    \vspace{3mm}
\end{table}

\clearpage

\section{Additional Tables}

\begin{table}[htbp]\centering
\caption{LMU and Trade - Figure 1}
\label{tab:lmu_trade}
\begin{tabular}{lcccccc}
\toprule
 & (1) & (2) & (3) & (4) & (5) & (6) \\
Variables & 1860---1873 & 1860---1885 & 1860---1913 & 1860---1873 & 1860---1885 & 1860---1913 \\
\midrule
lmu & 0.104 & -0.062 & -0.457*** & 0.274** & 0.117 & -0.183 \\
 & (0.078) & (0.116) & (0.167) & (0.111) & (0.136) & (0.183) \\

gs & -0.226*** & -0.182*** & -0.054 & -0.247*** & -0.190*** & -0.068 \\
 & (0.060) & (0.048) & (0.069) & (0.062) & (0.053) & (0.080) \\

other\_bimetallic 
&  &  &  & 0.150* & 0.190* & 0.312*** \\
 &  &  &  & (0.087) & (0.100) & (0.113) \\

silver &  &  &  & 0.276*** & 0.265** & 0.284* \\
 &  &  &  & (0.084) & (0.123) & (0.149) \\

paper &  &  &  & 0.133** & 0.004 & -0.018 \\
 &  &  &  & (0.060) & (0.044) & (0.077) \\

ta & 0.041 & 0.040 & 0.006 & 0.040 & 0.036 & 0.002 \\
 & (0.043) & (0.041) & (0.041) & (0.042) & (0.041) & (0.041) \\

\midrule
Exporter-year FE & Yes & Yes & Yes & Yes & Yes & Yes \\
Importer-year FE & Yes & Yes & Yes & Yes & Yes & Yes \\
Exp-imp FE & Yes & Yes & Yes & Yes & Yes & Yes \\
\midrule
Observations & 5,740 & 12,359 & 33,748 & 5,740 & 12,359 & 33,748 \\
\bottomrule
\end{tabular}

\vspace{3mm}
    \begin{minipage}{0.95\textwidth}
        \footnotesize \textbf{Notes}: Standard errors (in parentheses) are clustered at the directional country pair (i.e. exporter---importer). The dependent variable is bilateral trade from TRADHIST database. All specifications include exporter-year, importer-year, and exporter-importer fixed effects. *** $p<0.01$, ** $p<0.05$, * $p<0.1$.
    \end{minipage}
\end{table}
\clearpage


\begin{table}[htbp]\centering
\caption{LMU and Trade -- Robustness Checks, 1860--1873}
\label{tab:rob_lmu_trade_1860_1873}
\scriptsize
\setlength{\tabcolsep}{3pt}
\renewcommand{\arraystretch}{1.05}

\begin{tabular}{L*{8}{C}}
\toprule
 & (1) & (2) & (3) & (4) & (5) & (6) & (7) & (8) \\
\addlinespace[5pt]
Variables 
& \colhead{Wars}
& \colhead{Alliances}
& \colhead{Dom.\\trade}
& \colhead{gs $\neq$\\lmu}
& \colhead{ppml fe\\bias}
& \colhead{Three-way\\clustering}
& \colhead{Flandreau\\(2004)}
& \colhead{1855--1873} \\
\midrule

lmu 
& 0.271** & 0.274** & 0.361*** & 0.274** & 0.293** & 0.274*** & 0.262** & 0.379*** \\
& (0.110) & (0.111) & (0.123) & (0.111) & (0.140) & (0.106) & (0.110) & (0.136) \\

other\_bimetallic 
& 0.169** & 0.151* & 0.435*** & 0.150* & 0.160 & 0.150*** & 0.193** & 0.281** \\
& (0.086) & (0.087) & (0.111) & (0.087) & (0.106) & (0.050) & (0.089) & (0.118) \\

silver 
& 0.247*** & 0.274*** & 0.048 & 0.276*** & 0.266*** & 0.276*** & 0.311*** & 0.332*** \\
& (0.079) & (0.085) & (0.069) & (0.084) & (0.103) & (0.053) & (0.103) & (0.086) \\

paper 
& 0.121** & 0.133** & 0.008 & 0.133** & 0.131* & 0.133*** &  & 0.055 \\
& (0.059) & (0.060) & (0.064) & (0.060) & (0.073) & (0.042) &  & (0.067) \\

gs 
& -0.255*** & -0.245*** & -0.125** &  & -0.255*** & -0.247*** & -0.200*** & -0.319*** \\
& (0.064) & (0.063) & (0.062) &  & (0.094) & (0.092) & (0.068) & (0.065) \\

ta 
& 0.032 & 0.040 & 0.151*** & 0.040 & 0.050 & 0.040 & 0.049 & 0.094** \\
& (0.042) & (0.042) & (0.032) & (0.042) & (0.056) & (0.051) & (0.042) & (0.039) \\

war 
& -0.632*** &  &  &  &  &  &  &  \\
& (0.084) &  &  &  &  &  &  &  \\

alliance 
&  & 0.031 &  &  &  &  &  &  \\
&  & (0.118) &  &  &  &  &  &  \\

gs\_nolmu 
&  &  &  & -0.247*** &  &  &  &  \\
&  &  &  & (0.062) &  &  &  &  \\

Flandreau2004 
&  &  &  &  &  &  & 0.037 &  \\
&  &  &  &  &  &  & (0.051) &  \\

\midrule
Exporter-year FE 
& Yes & Yes & Yes & Yes & Yes & Yes & Yes & Yes \\
Importer-year FE 
& Yes & Yes & Yes & Yes & Yes & Yes & Yes & Yes \\
Exp-imp FE 
& Yes & Yes & Yes & Yes & Yes & Yes & Yes & Yes \\
\midrule
Observations 
& 5,740 & 5,740 & 6,027 & 5,740 & 5,740 & 5,740 & 5,740 & 7,494 \\
\bottomrule
\end{tabular}

\vspace{0.45cm}

\begin{tabular}{L*{8}{C}}
\toprule
 & (9) & (10) & (11) & (12) & (13) & (14) & (15) & (16) \\
\addlinespace[5pt]
Variables 
& \colhead{Excluding\\Italy}
& \colhead{López-Córdova\\and Meissner\\(2003)}
& \colhead{Timini\\(2023)}
& \colhead{Timini\\(2018)}
& \colhead{Europe\\only}
& \colhead{Europe w/o\\Germany}
& \colhead{Europe\\+ US}
& \colhead{RICardo} \\
\midrule

lmu 
& 0.211* & 0.350*** & 0.269** & 0.208* & 0.232** & 0.294*** & 0.406*** & 0.288** \\
& (0.117) & (0.123) & (0.111) & (0.107) & (0.108) & (0.111) & (0.122) & (0.119) \\

other\_bimetallic 
& 0.140 & 0.212** & 0.150* & 0.065 & 0.077 & 0.129 & 0.258** & 0.086 \\
& (0.092) & (0.100) & (0.087) & (0.086) & (0.088) & (0.088) & (0.101) & (0.127) \\

silver 
& 0.280*** & 0.281*** & 0.275*** & 0.452*** & 0.375*** & 0.255*** & 0.381*** & 0.432*** \\
& (0.084) & (0.086) & (0.084) & (0.107) & (0.089) & (0.089) & (0.088) & (0.097) \\

paper 
& 0.158** & 0.146** & 0.133** & 0.145 & 0.191 & 0.213* & 0.168** & 0.194 \\
& (0.069) & (0.062) & (0.060) & (0.172) & (0.120) & (0.124) & (0.082) & (0.196) \\

gs 
& -0.248*** & -0.251*** & -0.246*** & -0.256*** & -0.266*** & 0.174** & -0.253*** & 0.154 \\
& (0.063) & (0.062) & (0.061) & (0.067) & (0.066) & (0.069) & (0.069) & (0.236) \\

ta 
& 0.010 & 0.030 & 0.037 & 0.068 & 0.020 & 0.087** & 0.020 & -0.026 \\
& (0.048) & (0.043) & (0.042) & (0.042) & (0.046) & (0.037) & (0.050) & (0.048) \\

\midrule
Exporter-year FE 
& Yes & Yes & Yes & Yes & Yes & Yes & Yes & Yes \\
Importer-year FE 
& Yes & Yes & Yes & Yes & Yes & Yes & Yes & Yes \\
Exporter-importer FE 
& Yes & Yes & Yes & Yes & Yes & Yes & Yes & Yes \\
\midrule
Observations 
& 5,309 & 4,888 & 5,362 & 1,865 & 2,194 & 1,909 & 2,511 & 3,179 \\
\bottomrule
\end{tabular}

\vspace{3mm}
\begin{minipage}{0.95\textwidth}
\footnotesize
\textbf{Notes}: Standard errors in parentheses are clustered at the directional country-pair level (i.e. exporter---importer), except in Column (6), where standard errors are clustered by exporter, importer, and year. The dependent variable is bilateral trade from TRADHIST database, except in Column (16), where the RICardo import-flow measure is used. All specifications include exporter-year, importer-year, and exporter-importer fixed effects. Column (3) includes domestic trade and a time-varying international-border indicator. *** $p<0.01$, ** $p<0.05$, * $p<0.1$.
\end{minipage}
\end{table}
\clearpage
\textbf{Interpretation of monetary standard controls.} As discussed in Section~4.1, the specification is designed to identify the LMU coefficient. The remaining indicators identifying other monetary standards are included primarily to distinguish pairs sharing other monetary arrangements from pairs without a common arrangement, thereby defining the appropriate control group for the LMU. Their coefficients are not intended as separate causal estimates of the effects of the gold, silver, or bimetallic regimes (other than the LMU) and should instead be interpreted as conditional associations. For some regimes, these associations may also be sensitive to the limited identifying variation available in particular samples.

First, the coefficient on the gold standard dummy for the period 1860--1913 is close to zero. This is consistent with the pattern documented by \citet{Badia-MiroCamposTimini:25} when using three-way fixed effects and excluding domestic trade, and can be interpreted following the logic in \citet{CamposEtAl2026WTO}. In a similar setting, the authors argue that, as GATT/WTO membership becomes increasingly widespread by the end of the 20th century and beginning of the 21st century, estimates based only on international trade flows rely on a progressively smaller reference group of non-members. Once exporter-year, importer-year, and exporter-importer fixed effects are included, this leaves limited residual variation to identify the coefficient of interest. A similar concern applies here. As the sample approaches 1913, the gold standard becomes increasingly widespread. When domestic trade is not included, identification relies only on variation across international country pairs. As more countries adopt gold, the reference group of non-gold or mixed-regime pairs shrinks, leaving less variation to identify the gold standard coefficient. In line with their findings, when I estimate a specification (not reported in this paper) including domestic trade for the full sample 1860--1913, the coefficient on the gold standard turns positive and significant ($\beta=0.08$, s.e. $=0.04$). 

Second, when the sample is restricted to 1860--1873, our main period of interest, the gold standard coefficient is negative, as in \citet{Timini:23}. This estimate should not be interpreted as evidence of a negative causal effect of the gold standard on trade. In this short sample, the coefficient is identified by a small number of transitions into gold, whose timing is not always clear-cut. Germany is particularly important: Officer codes it as being on gold from 1871, although 1871--1873 are more appropriately regarded as transition years. The coefficient is therefore sensitive to the classification of a few country-year observations and becomes positive and significant when Germany is excluded from the sample (Appendix Table~\ref{tab:rob_lmu_trade_1860_1873}, Column 14). Similar concerns about the identification of a separate gold standard effect over this period may explain why \citet{AccominottiFlandreau2008} and \citet{Lampe:09} do not include a gold standard dummy in their analyses of the same years.

Third, before 1873, the LMU point estimate is roughly twice as large as the coefficient on other bimetallic standards, suggesting that the trade effect may reflect not only shared bimetallism but also the additional institutional content of a unified coinage framework. This interpretation is necessarily tentative, however, since the coefficient on other bimetallic standards is imprecisely estimated and based on very few observations. 

Fourth, the positive and statistically significant coefficient on the silver standard dummy is also consistent with previous evidence for Japan \citep{MitchenerShizumeWeidenmier:10}.

\clearpage

\section{Theoretical Appendix}
\label{sec:theory_appendix}
In this Appedix, I describe a simplified version of the model used in Section~\ref{sec:core_per}, with labor as the onyl factor of production.

Let $X_{ij}\ge 0$ denote the value of trade flows from country $i$ (exporter) to country $j$ (importer). The case $i=j$ denotes intra-national (domestic) trade and $i\neq j$ international trade. In a standard structural gravity system, bilateral flows satisfy
\begin{equation}
  X_{ij} \;=\; \frac{Y_i\,E_j}{Y}\left(\frac{\tau_{ij}}{\Omega_i\,\Pi_j}\right)^{-\theta},
  \label{eq: gravity_theory}
\end{equation}
where $Y_i \equiv \sum_j X_{ij}$ is production (income) in $i$, $E_j \equiv \sum_i X_{ij}$ is expenditure in $j$, and $Y \equiv \sum_i Y_i = \sum_j E_j$ is world income. Structural gravity further imposes the multilateral resistance (MR) conditions
\begin{equation}
  \Omega_i^{-\theta} \;=\; \sum_j \left(\frac{\tau_{ij}}{\Pi_j}\right)^{-\theta}\frac{E_j}{Y},
  \label{eq: outward_MR}
\end{equation}
\begin{equation}
  \Pi_j^{-\theta} \;=\; \sum_i \left(\frac{\tau_{ij}}{\Omega_i}\right)^{-\theta}\frac{Y_i}{Y},
  \label{eq: inward_MR}
\end{equation}
where $\Omega_i$ is the outward MR term (exporter $i$’s access to destination markets) and $\Pi_j$ is the inward MR term (the extent of competitive supply facing importers in $j$). Higher trade costs $\tau_{ij}$ (tariffs and non-tariff barriers, geography, culture, etc.) reduce $X_{ij}$; MR terms summarize the general-equilibrium influence of all partners on each bilateral flow. So far, this part is a brief overview of the information previously outlined in the main text, from Equation~\ref{eq: gravity theory} to  Equation~\ref{eq: inward MR}.

Define import shares
\begin{equation}
  \lambda_{ij} \;\equiv\; \frac{X_{ij}}{E_j} \;=\; \frac{Y_i}{Y}\left(\frac{\tau_{ij}}{\Omega_i\,\Pi_j}\right)^{-\theta},
  \label{eq: share_def}
\end{equation}
so that $\sum_i \lambda_{ij}=1$ and $\lambda_{ii}$ measures $i$’s home share.

\paragraph{Hat algebra.} For any variable $x$, let $\widehat{x}\equiv x'/x$ denote the change between a counterfactual ($'$) and the benchmark. From \eqref{eq: gravity theory},
\begin{equation}
  \widehat{X}_{ij} \;=\; \widehat{Y}_i\;\widehat{E}_j\;\widehat{\tau}_{ij}^{-\theta}\;\widehat{\Omega}_i^{\theta}\;\widehat{\Pi}_j^{\theta}.
  \label{eq:X_hat}
\end{equation}

\paragraph{Model structure and the share sufficient statistic.} In the workhorse quantitative models (Armington, Eaton–Kortum, Melitz, etc.), with inelastic labor as the only factor, the change in import shares depends only on wages and bilateral trade costs. The standard Dekle–Eaton–Kortum share change is
\begin{equation}
  \widehat{\lambda}_{ij}
  \;=\;
  \frac{\left(\widehat{w}_i\right)^{-\theta}\;\widehat{\tau}_{ij}^{-\theta}}
       {\sum\limits_{k}\lambda_{kj}\,\left(\widehat{w}_k\right)^{-\theta}\;\widehat{\tau}_{kj}^{-\theta}}.
  \label{eq:lambda_hat_DEK}
\end{equation}

\paragraph{Market clearing and the wage fixed point.} With labor as the only factor, $Y_i=w_i L_i$ and $\widehat{Y}_i=\widehat{w}_i$ when $L_i$ is fixed. Market clearing implies
\begin{align}
  \widehat{w}_i
  \;=\;
  \frac{1}{Y_i}\sum_{j} X_{ij}' 
  \;=\;
  \frac{1}{Y_i}\sum_{j} \lambda_{ij}'\,E_j'
  \;=\;
  \frac{1}{Y_i}\sum_{j} \lambda_{ij}\,\widehat{\lambda}_{ij}\,E_j',
  \label{eq:w_clearing_level}
\end{align}
where we used $X_{ij}'=\lambda_{ij}'E_j'$ and $\lambda_{ij}'=\lambda_{ij}\widehat{\lambda}_{ij}$.

In general, expenditure does not equal production because there are trade deficits. A trade deficit is defined by $E_j = Y_j + D_j$ and $E_j' = Y_j \hat{Y}_j + D_j \hat{D}_j$. There are two alternative assumptions that are commonly made to deal with the evolution of trade deficits. The first consists in the deficit to be ``additive''. This means that the deficit remains constant and $\hat{D}_j = 1$. The second consists in the deficit to be ``multiplicative''. This means that the deficit evolves in proportion to GDP, so that $\hat{D}_j = \hat{Y}_j$. 
 In the first case, $E_j' = Y_j \hat{Y}_j + D_j = Y_j \hat{w}_j + D_j$, and in the second case $E_j' = E_j \hat{Y}_j = E_j \hat{w}_j$.
 
In practical terms, the two assumptions are similar for counterfactual results, as in most cases, numbers obtained with one or the other option tend to be very similar. For the purpose of this demonstration, let trade deficits evolve \emph{multiplicatively} with GDP, so $E_j'=E_j\,\widehat{w}_j$. Substituting \eqref{eq:lambda_hat_DEK} and $E_j'$ into \eqref{eq:w_clearing_level} gives the fixed-point system for wages:
\begin{equation}
  \widehat{w}_i
  \;=\;
  \frac{1}{Y_i}\sum_{j}
  \lambda_{ij}\,
  \frac{\left(\widehat{w}_i\right)^{-\theta}\,\widehat{\tau}_{ij}^{-\theta}}
       {\sum\limits_{k}\lambda_{kj}\,\left(\widehat{w}_k\right)^{-\theta}\,\widehat{\tau}_{kj}^{-\theta}}
  \;E_j\,\widehat{w}_j.
  \label{eq:wage_fp}
\end{equation}
Because of Walras’ Law, \eqref{eq:wage_fp} is homogeneous of degree zero in $\{\widehat{w}_i\}$ and requires a normalization (we keep the nominal world output constant across scenarios, as in Baier et al., 2019).

\paragraph{Recovering other variables.} Once $\{\widehat{w}_i\}$ are solved from \eqref{eq:wage_fp}, the remaining objects follow:
\begin{align}
  &\widehat{Y}_i=\widehat{w}_i, 
  &&\widehat{E}_i=\widehat{w}_i, \label{eq:YE_hats}\\[2pt]
  &\widehat{\lambda}_{ij} \text{ from \eqref{eq:lambda_hat_DEK}}, 
  &&\widehat{X}_{ij} \;=\; \widehat{\lambda}_{ij}\,\widehat{E}_j. \label{eq:X_via_lambda}
\end{align}
When needed, inward MR changes can be recovered directly from baseline import shares:
\begin{equation}
  \widehat{\Pi}_j^{-\theta}
  \;=\;
  \sum_{k}\lambda_{kj}\,\left(\widehat{w}_k\right)^{-\theta}\,\widehat{\tau}_{kj}^{-\theta}.
  \label{eq:Pi_hat}
\end{equation}
(An explicit closed-form for $\widehat{\Omega}_i$ is not required for computing \eqref{eq:X_via_lambda}.)

\paragraph{Welfare.} Under standard CES preferences used in structural gravity, welfare changes are given by the change in real income, $\widehat{G}_i = \widehat{E}_i / \widehat{P}_i$, with the price index change pinned down by the inward MR. Using the structure above and the multiplicative deficit assumption, one obtains the familiar sufficient statistic
\begin{equation}
  \widehat{G}_i
  \;=\;
  \widehat{\lambda}_{ii}^{-1/\theta}.
  \label{eq:welfare_formula}
\end{equation}

\end{document}